\documentclass[12pt,preprint]{aastex}

\def\gsim{\;\lower4pt\hbox{${\buildrel\displaystyle >\over\sim}$}\;}
\def\lsim{\;\lower4pt\hbox{${\buildrel\displaystyle <\over\sim}$}\;}

\shorttitle{Measuring Bias through Weak Lensing}
\shortauthors{Fan}

\begin{document}


\title{Measuring the Deviation from the Linear and Deterministic Bias \\ 
through Cosmic Gravitational Lensing Effects}


\author{Zuhui Fan\altaffilmark{1,2,3}}


\altaffiltext{1}{Department of Astronomy, Peking University, Beijing 100871, China}
\altaffiltext{2}{Beijing Astrophysical Center, Chinese Academy of Science and
Peking University, Beijing 100871, China}
\altaffiltext{3}{Department of Astronomy and Astrophysics, The University
of Chicago, 5640 South Ellis Avenue, Chicago, IL 60637}


\begin{abstract}
Since gravitational lensing effects directly probe inhomogeneities of 
dark matter, lensing-galaxy cross-correlations can provide us important information on
the relation between dark matter and galaxy distributions, i.e., the bias.
In this paper, we propose a method to measure the stochasticity/nonlinearity of the
galaxy bias through correlation studies of the cosmic shear and
galaxy number fluctuations. Specifically, we employ the aperture mass statistics
$M_{ap}$ to describe the cosmic shear. We divide the foreground galaxy 
redshift $z_f<z_s$ into several bins, where $z_s$ is the redshift of the source galaxies, 
and calculate the quantity $<M_{ap}N_g(z_f)>^2/< N_g^2(z_f)>$ for each redshift bin. 
Then the ratio of the summation of $<M_{ap}N_g(z_f)>^2/< N_g^2(z_f)>$ over the bins 
to $<M_{ap}^2>$ gives a measure of
the nonlinear/stochastic bias. Here $N_g(z_f)$ is the projected surface number density
fluctuation of foreground galaxies at redshift $z_f$, and $M_{ap}$ is the aperture mass
from the cosmic-shear analysis. We estimate that for a moderately deep weak-lensing 
survey with $z_s=1$, source galaxy surface number density
$n_b=30 \hbox { gal}/\hbox { arcmin}^2$ and a survey area of $25 \hbox { deg}^2$,
the effective $r$-parameter that represents the deviation from the
linear and deterministic bias is detectable in the angular range of
$1'$-$10'$ if $|r-1|\gsim 10\%$. For shallow, wide surveys such as the Sloan Digital
Sky Survey with $z_s=0.5$, $n_b=5 \hbox { gal}/\hbox { arcmin}^2$, 
and a survey area of $10^4 \hbox { deg}^2$,
a $10\%$ detection of $r$ is possible over the angular range $1'-100'$.
\end{abstract}




\section{Introduction}

Galaxy surveys have provided us abundant knowledge of the large-scale structures
of the universe. To extract cosmological information from galaxy surveys, however, it is
crucial to understand the bias, i.e., the relation between the galaxy distribution
and the underlying dark matter distribution (e.g., Strauss 1999 and references therein).
The simplest bias model assumes a constant (in scale) bias factor between the two-point
correlation of galaxies and that of dark matter: i.e.,
$$
\xi_{gg}(r)=b^2 \xi(r), \eqno (1)
$$
where $\xi_{gg}$ is the two-point correlation function of galaxies
and $\xi$ is the two-point auto-correlation of dark matter. The $b$-factor
is referred to as the bias factor, which can be a function
of the galaxy type, brightness, etc.. The above relation is further extended to
$$
\delta_g=b\delta, \eqno (2)
$$
where $\delta_g=\delta n/n$, the galaxy number density fluctuation,
and $\delta=\delta \rho/\rho$, the dark matter mass density fluctuation.
Note that relation (2) is more restrictive than relation (1) and is
referred to as the linear and deterministic bias relation. Under this assumption,
it is straightforward to get the dark matter clustering properties from those of
galaxies since only the amplitude of clustering needs to be adjusted. In reality,
the bias could be much more complex than this. The bias factor $b$ can be
scale-dependent, which would introduce complexities in converting the galaxy distribution
to the underlying dark matter distribution. Further more, since galaxy formation
is a complicated process, it is expected that there should be a certain level of nonlinearity
and stochasticity in the relative distribution between the dark matter and galaxies
(e.g., Dekel \& Lahav 1999). Indeed the deviations from the linear and deterministic bias
have been seen in numerical simulations (e.g., Dekel \& Lahav 1999; Blanton et al. 1999).
It is shown that older galaxies and the underlying dark matter 
are highly correlated, but younger galaxies are poorly correlated with 
dark matter (e.g., Blanton et al. 1999, 2000; Somerville et al. 2001)
 
As far as the second moments are concerned, the nonlinear/stochastic bias can
be described by the quantities $b$ and $r$ which are defined as
$$
b^2\equiv {\sigma_g^2\over \sigma^2}, \eqno (3)
$$
and
$$
r\equiv {<\delta_g\delta>\over \sigma_g \sigma}, \eqno (4)
$$
where $\sigma^2\equiv <\delta^2>$ and $\sigma_g^2\equiv <\delta_g^2>$.
In the case of linear and deterministic bias, $r=1$.
 
It has been proposed to measure the nonlinearity/stochasticity of the bias
using the information of redshift distortions of the galaxy distribution,
which are presumably caused by the underlying dark matter distribution
(Dekel \& Lahav 1999; Pen 1998). Tegmark and Bromley (1999) analyzed
galaxy correlations in the Las Campanas Redshift Survey (
Shectman et al. 1996; Bromley et al. 1998). They found that different
subpopulations of galaxies are not correlated perfectly with each other,
indicating that the galaxy densities and the underlying dark matter
density cannot be perfectly correlated, or in other words, the bias
is nonlinear/stochastic. They further showed that the deviation from the linear
and deterministic bias is significant for late-type galaxies, and the $r$ factor ranges
from $\sim 0.98$ for very early-type galaxies to $\sim 0.57$ for
late-type ones. However, Blanton (2000) pointed out that this large
stochasticity (small $r$) could be the result of selection effects, and 
his estimation of $r$ is $r \approx 0.95$. 
 
Gravitational lensing effects due to large-scale structures (cosmic shear)
have been becoming important probes for studying the dark matter distribution
in the universe (e.g., Blandford et al. 1991; Kaiser 1992; Schneider et al. 1998;
Mellier 1999; Bartelmann \& Schneider 2001). Although weak ($\sqrt {< \gamma^2>}$ is
on the order of $\sim 1\%$ on angular scales $ < 10' $, where $< \gamma^2>$
is the top-hat variance of the cosmic shear),
the cosmic-shear signals have been detected by different groups (e.g., Mellier et al. 2002
and the references there in), and useful constraints on $\sigma_8$ and $\Omega_0$
have been set up with these lensing analyses (e.g., Maoli et al. 2001).
Recently, galaxy-galaxy lensing has also been detected (Fisher et al. 2000;
McKay et al. 2001; Hoekstra et al. 2002), and this enables the statistical study of
properties of galactic halos and galaxy-mass correlations (e.g, Guzik \& Seljak 2001, 2002).
 
While the most direct gain from cosmic-shear analyses is the projected dark matter
distribution, or the projected power spectrum in Fourier domain (Bartelmann \& Schneider 1999;
van Waerbeke et al. 1999), the cross-correlation between the cosmic shear and the
galaxy distribution reflects the relative distribution
of dark matter and galaxies, i.e., the bias. The combination of this cross-correlation and
the galaxy autocorrelation has been used to measure the scale dependency of the $b$-factor
(van Waerbeke 1998; Hoekstra et al. 2001). Hoekstra et al. (2002) also made a first try
to measure the $r$-factor by using the cosmic shear-galaxy, galaxy-galaxy, and cosmic
shear-cosmic shear correlations. In their sample, the foreground galaxies occupy
a relatively narrow redshift range. On the other hand, the mass contribution to the
cosmic shear comes from a relatively broad redshift range. Therefore they need a conversion factor
to convert the direct measurement result to the $r$-factor (see eq.[17] in Hoekstra et al. 2002).
The conversion factor is model-dependent, and this adds complications to the interpretation
of the observational results.
 
With fast developments in observations, it will be possible to divide galaxies
according to their redshifts, which can be obtained at least photometrically. This
will enhance the power to extract cosmological information from lensing analyses.
 
In this paper, we propose to use foreground galaxies in several redshift bins with
$z_f<z_s$ where $z_s$ is the source redshift, to get a summation
of $<M_{ap}N_g(z_f)>^2/<N_g^2(z_f)>$ over the bins, which should give a good estimation of
$<M_{ap}^2>$ if the bias is linear and deterministic, i.e., $r=1$. In other words,
the ratio $[\Sigma_{z_f} <M_{ap}N_g(z_f)>^2/<N_g^2(z_f)>]/<M_{ap}^2>$ measures
the projected $r^2$-factor weighted by the power spectrum in different redshift bins.
Here $M_{ap}$ is the aperture mass used to describe cosmic shear signals, and
$N_g$ is the surface number density fluctuations of foreground galaxies.
This measurement is largely independent of cosmology and of specific types of power spectra, and
there is no conversion factor between $r$ and the direct measurement result.
 
The rest of the paper is organized as follows. Section 2 discusses the effects of
different binning of redshift of foreground galaxies on the $r$-estimation,
i.e., the intrinsic uncertainties.
In section 3 we estimate the signal-to-noise level of the measurement of $r$ for
various models. Section 4 includes discussions.

\section{Redshift Binning}

In this paper, the aperture-mass statistics are used in the cosmic-shear
analysis (e.g., Schneider et al. 1998). The aperture mass $M_{ap}(\theta_c)$ is
defined as the projected surface mass density filtered with a compensated
filter function (Schneider et al. 1998)
$$
M_{ap}(\theta_c)\equiv \int d^2 \vec {\theta} U(\theta,\theta_c)
\kappa (\vec {\theta}), \eqno (5)
$$
where $\kappa$ is the dimensionless surface mass density, $U$ is a compensated filter
function with $U=0$ for $\theta\ge \theta_c$. The quantity $M_{ap}$ is related to the observable
shear $\gamma$ through
$$
M_{ap}(\theta_c)=\int d^2 \vec {\theta} Q(\theta,\theta_c) \gamma_t(\vec \theta),
\eqno (6)
$$
where $\gamma_t$ is the tangential component of the shear $\gamma$,
and $Q(\theta,\theta_c)=(2/\theta^2)\int_0^{\theta} d\theta^{'} \theta^{'}
U(\theta^{'},\theta_c)-U(\theta,\theta_c)$.
 
The surface mass density $\kappa$ is associated with the density
perturbation along the line of sight by
$$
\kappa (\vec \theta) ={3\over 2}\bigg ({H_0\over c}\bigg )^2\Omega_0\int_0^{w_H}
dw g(w)f_K(w) {\delta(f_K(w)\vec \theta,w)\over a(w)}, \eqno (7)
$$
where $H_0$ is the Hubble constant, $\Omega_0$ is the current matter density parameter of
the universe, $c$ is the speed of light, $w$ is the comoving radial
distance, $w_H$ is the comoving radial distance to the horizon, $f_K$
is the comoving angular diameter distance, $a$ is the scale factor,
$\delta$ is the density perturbation, and $g$ is defined as
$$
g(w)=\int_w^{w_H} dw' p_w(w'){f_K(w'-w) \over f_K(w')}, \eqno (8)
$$
where $p_w(w)$ is the distance distribution of source galaxies, and
$p_w(w)dw=p_z(z)dz$, where $p_z(z)$ is their corresponding redshift-distribution.
We use the filter function of Schneider et al. (1998), which has the form $U(\theta,\theta_c)
=u(\theta/\theta_c) /\theta_c^2$ with $u(x)=0$ for $x>1$,
and $Q(\theta,\theta_c)=q(\theta/\theta_c)/\theta^2$, 
$$
u(x)={(l+2)^2\over \pi}(1-x^2)^l\bigg ({1\over l+2}-x^2 \bigg ),
$$
and
$$
q(x)={(1+l)(2+l)\over \pi}x^2(1-x^2)^l,
$$
with $l$ an integer.
 
Then in the Fourier domain, we have
$$
<M_{ap}^2(\theta_c)>={9\pi\over 2}\bigg ({H_0\over c}\bigg )^4 \Omega_0^2\int_0^{w_H}
dw {g^2(w)\over a^2(w)} \times \int ds\ s\ P\bigg ({s\over f_K(w)}, w\bigg )\ I_l^2(s\theta_c) ,
\eqno (9)
$$
where $P$ is the three-dimensional power spectrum of density fluctuations
at the time corresponding to $w$, and $I_l$ is defined as
$$
I_l(\eta)= \int_0^1 dx\ x\ u(x) J_0(x\eta) ={2^l\Gamma(l+3)\over\pi} \eta^{-(l+1)}
J_{l+3}(\eta).
$$
In the following we will use $l=1$, and then
$$
I(\eta)={2\Gamma(4)\over\pi} \eta^{-2} J_{4}(\eta).  \eqno (10)
$$
 
In comparison with the top-hat window function, the compensated filter function (10)
is very localized with a peak at $\eta \sim 4$, and therefore $<M_{ap}^2(\theta_c)>$
provides a good measurement on the power spectrum of projected density fluctuations
at scale $s\sim 4/\theta_c$ (Bartelmann \& Schneider 1999).
When the cross-correlation between $M_{ap}$ and the
foreground-galaxy distribution is considered, the narrowness of the redshift distribution
of foreground galaxies makes the cross-correlation signals come from a well-defined
cosmic time. Thus, in conjunction with $I^2$, the contributions of the mass distribution
to the cross-correlation peak at a well-defined scale $k \approx (4/\theta_c)/f_K(z_f)$,
where $z_f$ is the peak redshift of foreground galaxies and $f_K(z_f)$ is the angular
diameter distance to $z_f$.
This property has been used in studying the scale dependence of $r/b$ through the 
measurement of the lensing-foreground-galaxy cross-correlation and the
auto-correlation of foreground-galaxy distributions (Schneider 1998, van Waerbeke 1998).  
In the analysis presented in this paper, we also use this property as 
explained in the following.

The projected distribution of foreground galaxies is described by
$$
N_g(\vec \theta)={N(\vec \theta)-\bar N\over \bar N}, \eqno (11)
$$
where $N(\vec \theta)$ is the surface number density of galaxies in the direction $\vec \theta$,
and $\bar N$ is the mean surface number density of galaxies. Then the cross-correlation
of $M_{ap}$ and $N_g$ is
$$
<M_{ap}(\theta_c) N_g(\theta_c)>=3\pi\bigg ( {H_0\over c}\bigg )^2 \Omega_0\ b\ r
\int dw {p_f(w)g(w)\over a(w) f_K(w)} \int ds\ s\ P\bigg ( {s\over f_K(w)}, w\bigg )
I^2 (s\theta_c), \eqno (12)
$$
and the auto correlation of $N_g$ is
$$
<N_g^2(\theta_c)>=2\pi b^2 \int dw {p_f^2(w)\over f_K^2(w)}
\int ds\ s\ P\bigg ( {s\over f_K(w)}, w\bigg ) I^2 (s\theta_c), \eqno (13)
$$
where $p_f$ is the foreground-galaxy distribution, $b$ and $r$ are the bias
factors discussed in the section 1, and $r=1$ if the bias is
linear and deterministic.
 
As stated before, if $p_f$ is highly peaked at $z_f$, then
$$
<M_{ap}(\theta_c) N_g(\theta_c)>\  \propto 3\pi\bigg ( {H_0\over c}\bigg )^2 \Omega_0\ b\ r
{g(z_f)\over a(z_f) f_K(z_f)} P(k, z_f),  \eqno (14)
$$
and
$$
<N_g^2(\theta_c)> \  \propto 2\pi b^2 {p_f(z_f) (dz/dw)_{z_f} \over f_K^2(z_f)} P(k, z_f), \eqno (15)$$
with $k \approx (4/\theta_c)/f_K(z_f)$. Notice that in equations (14) and (15), we use $z_f$
instead of $w$ to label the time, and the formulation for $g$ and $f_K$
should be changed accordingly in the numerical calculations.
Then
$$
{[<M_{ap}(\theta_c) N_g(\theta_c)>]^2\over <N_g^2(\theta_c)> } \times p_f(z_f)
\ \propto r^2 \bigg \{ {9\pi\over 2} \bigg ( {H_0\over c}\bigg )^4 \Omega_0^2
{g^2(z_f)\over a^2(z_f)} \bigg ({dw\over dz}\bigg )_{z_f} P(k, z_f) \bigg \}.  \eqno (16)
$$
 
From equation (9), it is seen that
$$
<M_{ap}^2(\theta_c)> \ \propto \bigg [ {9\pi\over 2} \bigg ( {H_0\over c}\bigg )^4 \Omega_0^2
\int dz {g^2(z)\over a^2(z)} \bigg ({dw\over dz}\bigg )_{z} P(k, z) \bigg ],
$$
with $k \approx (4/\theta_c)/f_K(z)$. Therefore if the foreground galaxies are divided into
several redshift bins with $z_f\le z_s$, and
$[<M_{ap}(\theta_c) N_g(\theta_c)>]^2/ <N_g^2(\theta_c)> $ is calculated for each bin,
the summation of $[<M_{ap}(\theta_c) N_g(\theta_c)>]^2/ <N_g^2(\theta_c)>  \times p_f(z_f)$
over the bins would give rise to an estimation of $<M_{ap}^2(\theta_c)>$ if $r=1$. With
$r\neq 1$, then we have
$$
\Sigma_{z_f} \bigg \{ {[<M_{ap}(\theta_c) N_g(\theta_c,z_f)>]^2\over <N_g^2(\theta_c,z_f)> }  p_f(z_f) \Delta z_f \bigg \} \approx r^2 <M_{ap}^2(\theta_c)>,
$$
or
$$
r^2 \approx {\Sigma_{z_f} \bigg \{ [<M_{ap}(\theta_c) N_g(\theta_c,z_f)>]^2/ <N_g^2(\theta_c,z_f)>
 \times p_f(z_f) \Delta z_f \bigg \} \over <M_{ap}^2(\theta_c)> }. \eqno (17)
$$
Here we have explicitly written down the redshift dependence of $N_g$. The approximate sign
" $\approx$ " means that when equation (17) is used in estimating $r$, there are intrinsic 
uncertainties depending on the redshift binning. The binning effects are discussed below.
In equation (17), it is assumed that $r$ is independent of scales. 
In general, $r$ is scale-dependent, and then equation (17) provides a measurement of the 
power-spectrum-weighted $r$-factor. In addition, theoretical studies and numerical 
simulations showed that the $r$ factor evolves with time, i.e., it should be
a function of redshift $z$ (e.g., Tegmark \& Peebles 1998; Blanton et al. 2000;
Somerville et al. 2001). Then, the method proposed here really measures the weighted
averaged $r$ over the redshift range up to $z_s$, the source redshift. 
 
With $r=1$, we expect the right hand side of equation (17) 
(denoted as $r^2_{es}$) to be close to $1$
if the binning is fine enough. On the other hand, it is observationally impractical
if a very fine binning is required. In the following, we study how $r^2_{es}$ deviates
from $1$ with different binning in the case of linear and deterministic bias, i.e., $r=1$.
This gives us an evaluation on the intrinsic uncertainty when we use equation (17) to measure
the $r$-factor.
 
We assume all source galaxies are located at redshift $z_s$. The foreground-galaxy distribution
is written as
\begin{eqnarray}
&&\qquad\qquad p_f(z_f)={1\over z_2-z_1} \quad \hbox {for} \quad z_1\le z_f \le z_2 
\nonumber \\ &&\qquad\qquad\qquad\quad
=0 \qquad \quad \quad \hbox {otherwise} \qquad\qquad\qquad\qquad\qquad\qquad\qquad\qquad\qquad 
(18) \nonumber
\end{eqnarray}
Equal binning is applied in the calculations, i.e., $(z_2-z_1)=z_s/N_{bin}$ where
$N_{bin}$ is the number of bins used.  A nonlinear power spectrum in the form of
Peacock and Dodds (1996) is adopted.
 
Figure 1 presents the integrand of $<M_{ap}^2(\theta_c)>$ [denoted as $m_{ap}(z)$, see eq. (9)
but with the integrated variable changed from $w$ to $z$] with $z_s=1$ versus redshift $z$.
Figure 1a is for a $\Lambda$-dominated cold dark matter ($\Lambda$CDM) 
model with $\Omega_0=0.3$, $\Omega_{\Lambda}=0.7$,
$H_0=67 \hbox{ km/s/Mpc}$, the shape parameter $\Gamma=0.2$, and $\sigma_8=0.93$.
Different curves correspond to $\theta_c=1', 3', 5'$, and $10'$, respectively.
Figure 1b is for the open CDM model (OCDM) with $\Omega_0=0.3$, $\Omega_{\Lambda}=0$,
$H_0=67 \hbox{ km/s/Mpc}$, $\Gamma=0.2$, and $\sigma_8=0.87$.
For comparison, we plot in Figure 1c the results for both models with $\theta_c=5'$.
It can be seen that the contributions to $<M_{ap}^2(\theta_c)>$ spread over a fairly
large range of $z$. For OCDM, the low redshift contributions are more prominent than
those of the $\Lambda$CDM model.
 
In Figure 2 we show $1-r^2_{es}$ (assuming linear and deterministic bias) with $z_s=1$
versus $\theta_c$ for the $\Lambda$CDM model with $N_{bin}=5, 10, \hbox{ and } 20$,
which corresponds to $\Delta z=z_2-z_1=0.2,\hbox{ } 0.1, \hbox{ and } 0.05$,
respectively. We see that with $\Delta z=0.2$, the deviation of $r^2_{es}$ from $1$ is as
high as about $10\%$, 
and $1-r^2_{es}$ reaches $\sim 7\%$ at $20'$. Thus with $N_{bin}=5$, the accuracy
is not good enough to give a sound estimation of $r$ by using equation (17). 
With $N_{bin}=10$ and $\Delta z=0.1$, $1-r^2_{es}< 3\%$ for the whole angular range 
we studied, and at
$\sim 20'$, $r^2_{es}$ deviates from $1$ by only less than $2\%$.
With finer binning of $\Delta z=0.05$, $1-r^2_{es}< 0.7\%$ for angular
scales up to $100'$ with $1-r^2_{es}< 0.5\%$ around $20'$.
Therefore, with $\Delta z=0.05$, the intrinsic uncertainty of equation (17) is tiny, but
observationally, it may be difficult to precisely (with $0.05$ accuracy) measure
the redshift for a large number of foreground galaxies especially for those with
redshift close to $1$, since most likely only photometric redshifts will be available
for high-redshift galaxies. Considering both the intrinsic uncertainties and the observational
realities,  it is therefore optimal to use $N_{bin}=10$ ($\Delta z=0.1$) in the case of $z_s=1$.
 
Similar curves for the OCDM model are shown in Figure 3. It is seen that $1-r^2_{es}$ is
larger for the OCDM model than the corresponding $1-r^2_{es}$ for the $\Lambda$CDM model.
This is because as seen in Figure 1c, low-redshift contributions to $<M_{ap}^2>$ are
more significant in the OCDM model than in the $\Lambda$CDM model, and therefore for
OCDM, relatively finer binning is needed in low redshifts in order to reach the same $1-r^2_{es}$
as that of $\Lambda$CDM. Still, however, with $N_{bin}=10$ ($\Delta z=0.1$),
$1-r^2_{es}$ is small enough
with $1-r^2_{es} < 3.5\%$ at $20'$, and $1-r^2_{es} \le 4\%$ for the whole
angular range up to $100'$.
 
In Figure 4 we plot $1-r^2_{es}$ for the $\Lambda$CDM model ({\it Solid lines}) 
and OCDM model ({\it dashed lines})
with $z_s=1.5$. The two sets of curves correspond, respectively, to
$N_{bin}=10$ ($\Delta z=0.15$) and $N_{bin}=15$ ($\Delta z=0.1$).
For $N_{bin}=10$ ($\Delta z=0.15$), $1-r^2_{es}<3.5\% \hbox{ and } 5\%$ for the $\Lambda$CDM model
and OCDM model, respectively. At $20 \hbox{ arcmin}$, $1-r^2_{es}<2.5\% \hbox{ and } 4\%$ for
the respective two models. For $N_{bin}=15$ ($\Delta z=0.1$), the deviations of $r^2_{es}$ from
$1$ are, respectively, less than $1.5\%$ and $2\%$ for the two models for the whole angular
range considered. Thus for $z_s=1.5$, $N_{bin}=15$ ($\Delta z=0.1$) guarantees that
the intrinsic uncertainties in terms of $r$ estimation with equation (17) are small. Even with
$N_{bin}=10$ ($\Delta z=0.15$), the intrinsic uncertainties are insignificant.
 
Figure 5 shows the curves for $z_s=0.5$. The results for each of the two models
with $N_{bin}=5$ ($\Delta z=0.1$) and $N_{bin}=10$ ($\Delta z=0.05$) are shown in the plot.
For $N_{bin}=5$ and $\Delta z=0.1$, the deviations are relatively large, with
$1-r^2_{es}$ as high as $8\%$ and $10\%$ for the $\Lambda$CDM model and OCDM model, respectively.
With $N_{bin}=10$ and $\Delta z=0.05$, the deviations are less than $2\%$ and $3\%$ for
$\Lambda$CDM and OCDM, respectively, with $1-r^2_{es}<1.5\% \hbox { and } 2\%$ at
$20'$. Therefore, for $z_s=0.5$, it is necessary to go to $\Delta z=0.05$ in order
to control the intrinsic uncertainties in the $r$-estimation using equation (17). Fortunately,
for $z\le 0.5$, the redshifts of foreground galaxies can be obtained spectroscopically with
very high precision, and it is possible to bin them down to $\Delta z=0.05$.
 
From above analyses, we see that for source galaxies at $z_s$, in general $10$ bins
for the foreground galaxies would be enough to control the intrinsic uncertainties of 
equation (17)
to a reasonably low level. For $z_s\ge 1$, the intrinsic uncertainties can be well
controlled if the bin interval $\Delta z=0.1$ is used.




\section{Signal-to-Noise Ratio}

In this section, we discuss the signal-to-noise ratio of $r$ estimated with 
equation (17) considering the intrinsic ellipticity of source galaxies, 
Poisson noise of the foreground-galaxy distribution, and the cosmic variance.
 
If we denote 
$$ 
\{[<M_{ap}(\theta_c) N_g(\theta_c,z_f)>]^2/ <N_g^2(\theta_c,z_f)>
\} p_f(z_f) \Delta z_f
$$ 
as $M^2_{z_f}$, then equation (17) becomes
$$
r^2 \approx {\Sigma_{z_f} M^2_{z_f} \over <M_{ap}^2(\theta_c)> }. \eqno (19)
$$
 
To estimate the signal-to-noise ratio of $r$, we assume that the $M_{z_f}$ 
of different redshift bins are independent. This is a reasonable 
assumption with the bin intervals
we considered. Further we assume that the dispersion of $<N_g^2(\theta_c,z_f)>$ can be neglected
(van Waerbeke 1998). For one field, we then have
$$
{\sigma(M^2_{z_f})\over M^2_{z_f} }={2\sigma (<M_{ap}(\theta_c) N_g(\theta_c,z_f)>)
\over <M_{ap}(\theta_c) N_g(\theta_c,z_f)>}, \eqno (20)
$$
and
$$
\sigma^2 (\Sigma_{z_f} M^2_{z_f})=\Sigma_{z_f} \sigma^2(M^2_{z_f}). \eqno (21)
$$
 
If $\Sigma_{z_f}(M^2_{z_f})$ and $<M^2_{ap}(\theta_c)>$ are independent, then
$$
{\sigma(r^2)\over r^2}=\bigg \{ {\sigma^2(\Sigma_{z_f} M^2_{z_f}) \over (\Sigma_{z_f} M^2_{z_f})^2}+{\sigma^2(<M^2_{ap}(\theta_c)>)\over [<M^2_{ap}(\theta_c)>]^2} \bigg \}^{1/2}. \eqno (22)
$$
 
However, it is likely that $\Sigma_{z_f}(M^2_{z_f})$ and 
$<M^2_{ap}(\theta_c)>$ are not independent, so then (Taylor 1982)
$$
{\sigma(r^2)\over r^2}\le \sigma_{max}={\sigma(\Sigma_{z_f} M^2_{z_f}) \over \Sigma_{z_f} M^2_{z_f}}+{\sigma(<M^2_{ap}(\theta_c)>)\over <M^2_{ap}(\theta_c)>}. \eqno (23)
$$
 
In the following, both equation (22) and $\sigma_{max}$ defined in 
equation (23) are calculated.
 
We use the estimation of van Waerbeke (1998) to calculate
$\sigma [<M_{ap}(\theta_c) N_g(\theta_c,z_f)>]$, i.e.,
$$
\sigma (<M_{ap}(\theta_c) N_g(\theta_c,z_f)>)=\bigg [<M_{ap}^2(\theta_c)>
+{G\sigma_{\epsilon}^2\over 2N_b}\bigg ]^{1/2}\bigg [<N_g^2(\theta_c,z_f)>+
{\tilde G\over N_f}\bigg ]^{1/2},
\eqno (24)
$$
where $N_b$ and $N_f$ are the numbers of source and foreground galaxies
in one field, respectively, $\sigma_{\epsilon}$ represents the intrinsic ellipticity of source galaxies,
$G=\pi \theta_c^2\int d^2\theta Q^2(\theta,\theta_c)=6/5$ if the
compensated filter defined previously with $l=1$ is used, and
$\tilde G=\pi \theta_c^2\int d^2\theta U^2(\theta,\theta_c)=G$.
Within each pair of square brackets of the right-hand side of equation 
(24), the first
term represents the uncertainty due to the cosmic variance.
For the aperture mass, the intrinsic ellipticity of background galaxies
contributes to the uncertainty through the second term of the first square
bracket. The second term in the second square bracket represents the statistical
uncertainty due to the finite number of foreground galaxies.
 
For $\sigma(<M^2_{ap}>)$, we have (Schneider et al. 1998)
$$
\sigma(<M^2_{ap}>)\approx \bigg [ \mu_4 <M_{ap}^2>+\bigg ({\sigma_{\epsilon}^2G\over
\sqrt {2} N_b}+\sqrt {2} <M_{ap}^2>\bigg )\bigg ]^{1/2}, \eqno (25)
$$
where $\mu_4$ is the kurtosis of $M_{ap}$ with $\mu_4=<M_{ap}^4>/(<M_{ap}^2>)^2-3$,
which is $0$ if $M_{ap}$ is Gaussian. On small scales, however, non-linear effects
are strong, and $M_{ap}$ is non-Gaussian. As shown in van Waerbeke (2002),
$<M_{ap}^4>/<M_{ap}^4>_{Gaussian}$ can reach about $7$ for 
$\theta_c \le 3'$ for
$\Lambda$CDM model. On the other hand, on small scales, we expect statistical
uncertainties to dominate over the cosmic variance, 
thus, neglecting non-Gaussianity
should not affect the error estimation significantly. In the following, we 
ignore the $\mu_4$ term in equation (25).
 
Figure 6 shows $\sigma(r^2)/r^2$ of one field with $z_s=1$, $N_{bin}=10$ ($\Delta z=0.1$),
$n_b=30\hbox{ gal}/\hbox { arcmin}^2$, $\sigma_{\epsilon}=0.2$ and
$n_f=5\hbox{ gal}/\hbox { arcmin}^2$, where $n_b$ and $n_f$ are the surface number
densities of background and foreground galaxies, respectively. The solid lines are
for the $\Lambda$CDM model, and the dashed lines are for the OCDM model. In each model,
the upper line corresponds to the result of $\sigma_{max}$ in equation (23) and the lower line
corresponds to that of equation (22). It is seen that at $\theta_c\le 3'$, 
the statistical
uncertainties are dominant, and $\sigma(r^2)/r^2$ is larger for the $\Lambda$CDM model.
At angular scales $\theta_c > 3'$, the two models have about the same
$\sigma(r^2)/r^2$, and it approaches a constant when $\theta_c > 10' $.
 
In Figure 7 we plot $\sigma(r^2)/r^2$ for a survey of $25\hbox { deg}^2$. The meanings of
different lines are the same as those in Figure 6. We see that $\sigma(r^2)/r^2 \le 20\%$
or $\sigma(r)/r=1/2\sigma(r^2)/r^2 \le 10\%$ for $\theta_c\le 10'$. In other
words, the deviation of $r$ from $1$ is detectable if it is larger than $10\%$ at
$\theta_c\le 10'$. Going deep in a survey, the number density of source
galaxies increases and the statistical uncertainties decrease. As we have seen,
however, on relatively large scales, the cosmic variance dominates, and thus a deeper
survey will not help to increase the precision on the determination of $r$ on those
scales. On the other hand, by increasing the survey area, $\sigma(r^2)/r^2$ decreases by a same
factor over the whole angular range. In Figure 8, we compare the effects on $\sigma(r^2)/r^2$ with
a deeper survey and with a wider survey for the $\Lambda$CDM model. The solid
line shows the result of equation (22) with 
$n_b=30\hbox{ gal}/\hbox { arcmin}^2$ and a survey
area of $25\hbox { deg}^2$. The dashed line is for $n_b=60\hbox{ gal}/\hbox { arcmin}^2$ and
a survey area of $25\hbox { deg}^2$. The dot-dashed line is for
$n_b=30\hbox{ gal}/\hbox { arcmin}^2$ and a survey area of
$100\hbox { deg}^2$. It is clearly seen that in terms of determining $r$, it is more
effective to go wide than to go deep in a lensing survey.
 
Figure 9 shows the results for $z_s=1.5$ and $N_{bin}=15$ ($\Delta z=0.1$).
The other parameters are the same as in Figure 7.
At small angular scales, ($\theta_c\le 4'$), $\sigma(r^2)/r^2$ is smaller
in the case of $z_s=1.5$ because of relatively large cosmic-shear signals 
compared $z_s=1$. A $10\%$ detection is reachable for 
$\theta_c\le 10'$ with a survey of $25\hbox { deg}^2$.
 
In the case of $z_s=0.5$, since the shear signals are very weak, the statistical
uncertainties are relatively very large. Figure 10 plots $\sigma(r^2)/r^2$ for
$n_b=30\hbox{ gal}/\hbox { arcmin}^2$, $N_{bin}=10$ ($\Delta z=0.05$)
and a survey area of $25\hbox { deg}^2$ for the two cosmological models.
The number density of foreground galaxies is
$n_f=5\hbox{ gal}/\hbox { arcmin}^2$. We see that over the whole angular range,
$\sigma(r^2)/r^2 >20\%$. The good strategy is then to do shallow, wide surveys.
For the Sloan Digital Sky Survey (SDSS), the survey area can be as large as $10^4\hbox { deg}^2$. The surface number
density of source galaxies is about a few (e.g., Mckay et al. 2001).
In Figure 11 we show $\sigma(r^2)/r^2$ for $n_b=5\hbox{ gal}/\hbox { arcmin}^2$ and a survey area of $10^4\hbox { deg}^2$. 
All the other parameters are the same as in Figure 10.  It is seen that 
with such a large survey area, a $10\%$ detection of $r$ is possible for
all the angular scales we considered ($1'-100'$).

We would like to emphasize that in this section we only aimed at giving
an estimation on the signal-to-noise ratio of $r$, i.e., to what level
the deviation of $r$ from $1$ is detectable. For that, we only calculated
the standard deviation of $r$. Since the statistics of $r$
are certainly non-Gaussian, and rather complicated, strictly speaking
we cannot draw conclusions on the significance level of a measured $r$
based on Gaussian statistics.
For example, if we get a $2\sigma$ measurement on $r$, the Gaussian probability that
$r$ is consistent with $r=1$ is $5\%$. For real $r$, the probability can be very different
depending on the statistics of $r$.
Nonetheless, we expect that our signal-to-noise ratio
should give a reasonable estimation of the detectability of $r$. 
Detailed statistical properties of $r$ can be best studied through numerical
simulations, which is our planned next step.   
In addition, in our estimation, we implicitly assumed that the uncertainties 
were small, 
so that the error propagation rules were applicable. From our calculations discussed above,
the uncertainties are indeed small on the angular range of less
than $ < 10'$ if the survey area is large enough. 






\section{Discussion}


In this paper, we studied the feasibility of extracting information on $r$ from
the cross-correlation between the foreground-galaxy distribution and cosmic gravitational
lensing effects under the availability of the redshift information for each foreground galaxy.  
We showed that for $z_s\sim 1$, the intrinsic uncertainty in the $r$-estimation can be
well controlled if the foreground galaxies are binned with $\Delta z=0.1$. This binning requirement
is within the accuracy range of photometrically determined redshifts, 
and is thus achievable. For shallow
surveys with $z_s\sim 0.5$, a fine binning with $\Delta z=0.05$ is necessary. This fine binning
is possible, since the redshifts of low-redshift galaxies can be obtained spectroscopically.

We further found that for a moderately deep lensing survey
with $z_s \sim 1$ and $n_b=30 \hbox{ gal}/\hbox { arcmin}^2$, a $10\%$ detection on $|r-1|$
is possible in the angular range of $1'-10'$ with a survey area of
$25 \hbox { deg}^2$. A deeper survey with $n_b=60 \hbox{ gal}/\hbox { arcmin}^2$
helps to reduce uncertainties only on small angular scales where the noise
due to the intrinsic ellipticity of source galaxies dominates.
The effective way to reduce the uncertainties over a large angular range is to increase
the survey area. For example, with a survey area of $100\hbox { deg}^2$,
a $10\%$ detection of $|r-1|$ can be made up to $\theta_c \approx 20'$.
 
For $z_s=0.5$, because of the relatively low cosmic shear
signal, the statistical noise from the intrinsic ellipticity of source galaxies
is much more dominant than in the case of $z_s\sim 1$. Thus a very large survey area
is needed in order to detect the deviation of $r$ from $1$. SDSS will eventually
have a survey area of about $10^4\hbox { deg}^2$, and will be very suitable to
be used to measure $r$ with the method we propose here.
 
For the $\Lambda$CDM model, the lensing effects on source galaxies with $z_s=1$ mainly
come from the matter distribution around redshift $z\sim 0.18-0.68$, with
a peak located at $z\sim 0.4$ for $\theta_c=1'$. The three-dimensional
wave-vector
$k$ corresponding to $z\sim 0.4$ and $\theta_c=1'$ is $k\sim 12
(\hbox { Mpc}h^{-1})^{-1}$. For $\theta_c=10'$, the contributions
to the lensing effects peak at $z\sim 0.29$, with significant amounts from 
$z\sim 0.1-0.57$. The corresponding peak scale is $k\sim 1.7(\hbox { Mpc}h^{-1})^{-1}$.
Thus, measuring $r$ in the angular scale range $1'-10'$ with $z_s=1$, we probe
the bias of structure distribution on scales of about $0.05\hbox { Mpc}h^{-1}$ to about
$1\hbox { Mpc}h^{-1}$ at redshift $z\sim 0.1-0.7$.
For $z_s=0.5$, the lensing effects are mainly from matter fluctuations in the redshift
range $z\sim 0.08-0.4$ for $\theta_c=1'-10'$. The corresponding
scale is $\sim 0.06\hbox { Mpc}h^{-1}$ to $\sim 0.6\hbox { Mpc}h^{-1}$.
 
In the above studies, we have assumed that source galaxies 
are all located at the
same redshift $z_s$. For future surveys, it is possible to bin source
galaxies according to their redshifts, and thus our analyses here will be directly
applicable. Currently, most surveys select source galaxies according to their
luminosities. Their redshift distribution is approximately in the form
$p_b(z)\propto z^2exp[-(z/z_0)^{\beta}]$ with $\beta\approx 1.5$. 
In this case, in order
to control well the intrinsic uncertainties in the $r$-estimation, the redshift
range of foreground galaxies has to be extended to about $<z>=1.5z_0$. With $z_0=0.7$,
$<z>\approx 1$.
 
With the fast growth of gravitational lensing surveys,
we expect, with the additional information on the galaxy distribution, that we can obtain
a great deal of knowledge on the process of galaxy formation, which will further
help us to extract cosmological information from other types of surveys.



\acknowledgments
We thank the referee for the constructive comments that helped us
further clarify our studies. This research was supported by the National 
Science Foundation of China, under Grant 10243006.

\clearpage



\begin{figure}
\plottwo{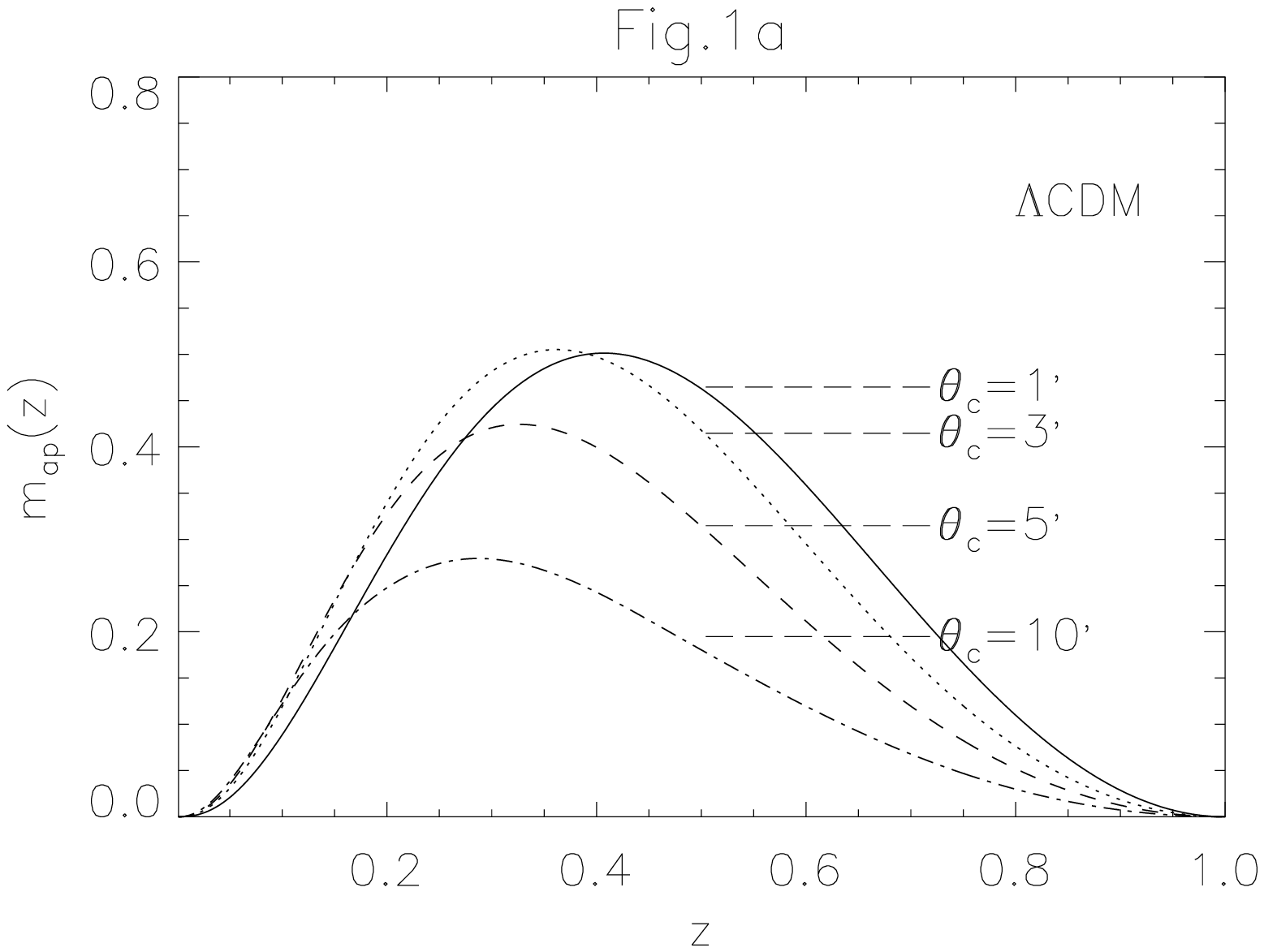}{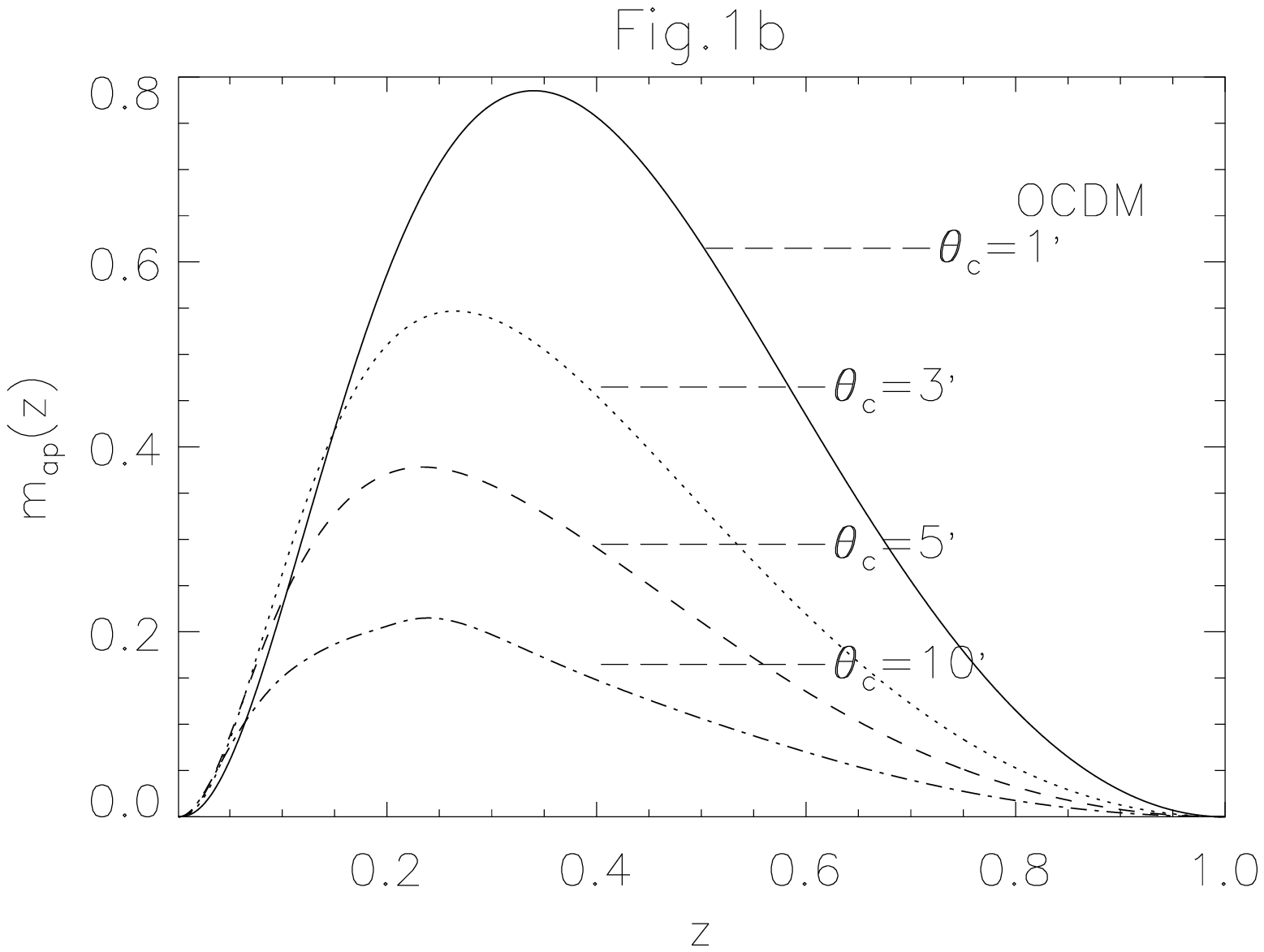}
\plotone{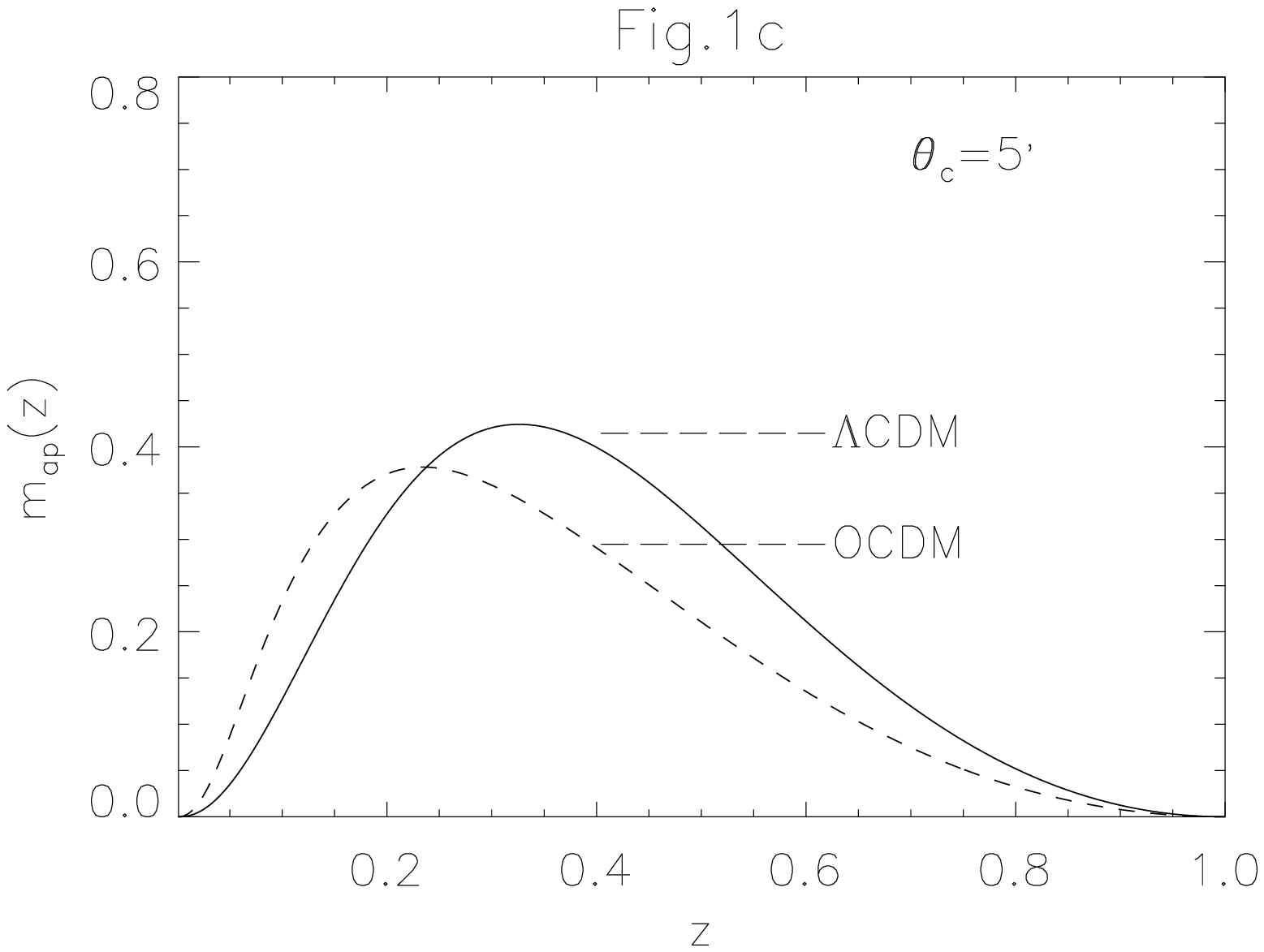}
\caption{Plot of $m_{ap}(z)$, the integrand of $<M_{ap}^2>$, vs. redshift $z$. (a) For
the $\Lambda$CDM model. The solid line is for $\theta_c=1'$, the dotted line is
for $\theta_c=3'$, the dashed line is for $\theta_c=5'$, and the dot-dashed
line is for $\theta_c=10'$. (b) For the OCDM model. (c) For $\theta_c=5'$,
where the solid line is for the $\Lambda$CDM model, and the dashed line is for
the OCDM model. \label{f1}}
\end{figure}

\begin{figure}
\plotone{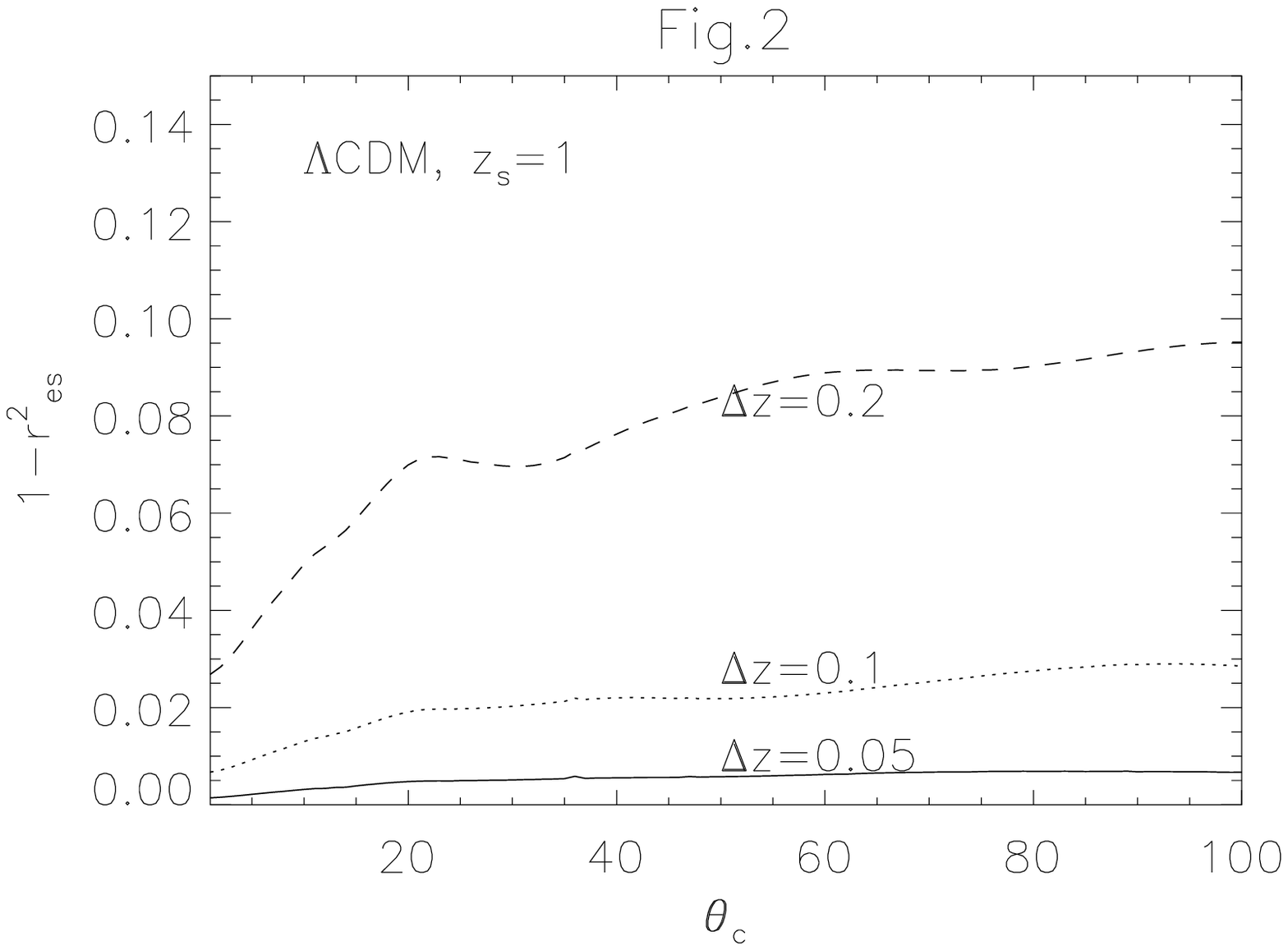}
\caption{Intrinsic uncertainty $1-r_{es}^2$ vs. the angular scale $\theta_c$ (arcmin) with
$z_s=1$ for the $\Lambda$CDM model. The solid line is for $\Delta z=0.05$, the dotted line
is for $\Delta z=0.1$, and the dashed line is for $\Delta z=0.2$.\label{f2}}
\end{figure}

\begin{figure}
\plotone{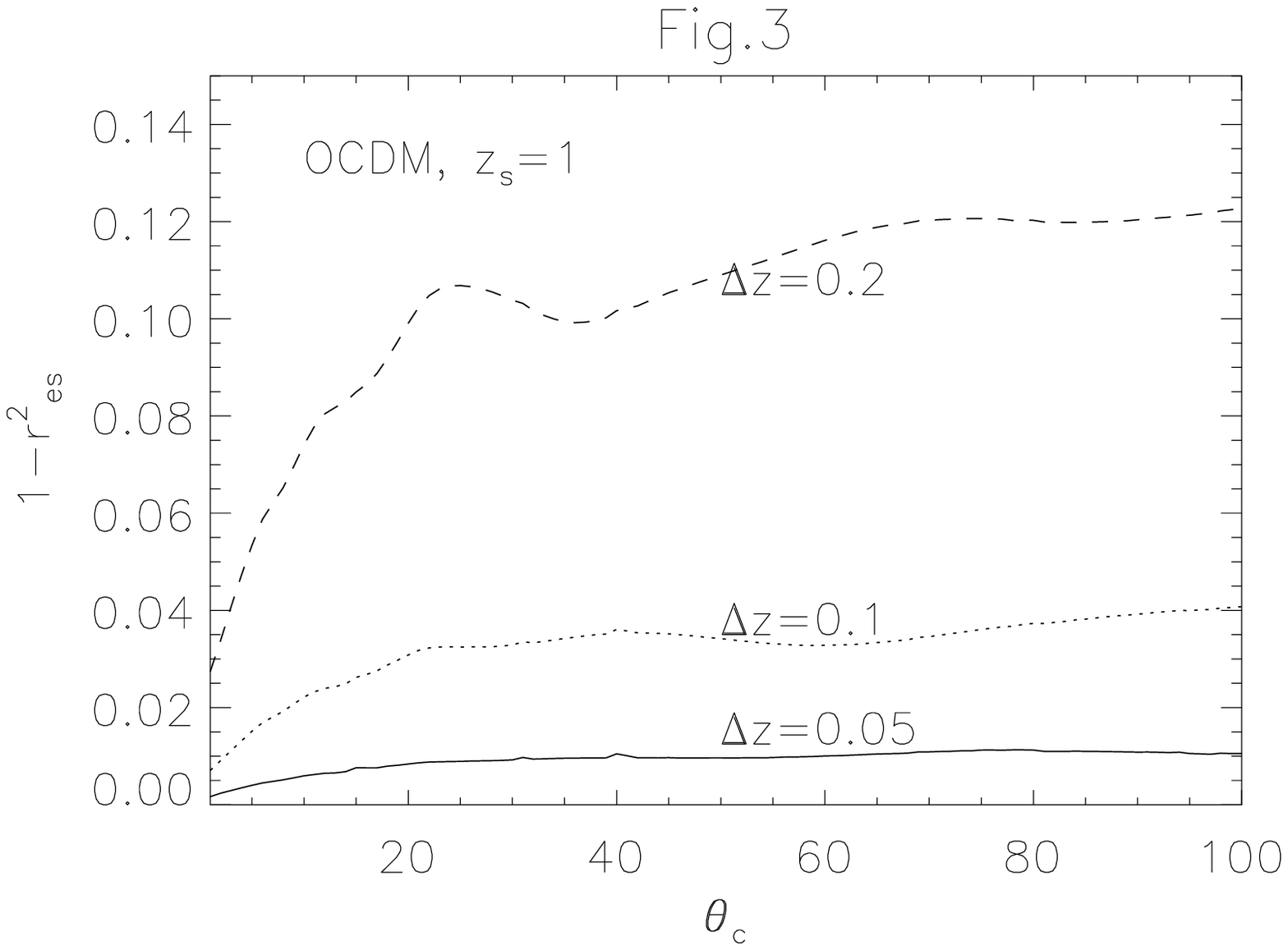}
\caption{Intrinsic uncertainty $1-r_{es}^2$ vs. the angular scale $\theta_c$ (arcmin) with
$z_s=1$ for the OCDM model. The solid line is for $\Delta z=0.05$, the dotted line
is for $\Delta z=0.1$, and the dashed line is for $\Delta z=0.2$.\label{f3}}
\end{figure}

\begin{figure}
\plotone{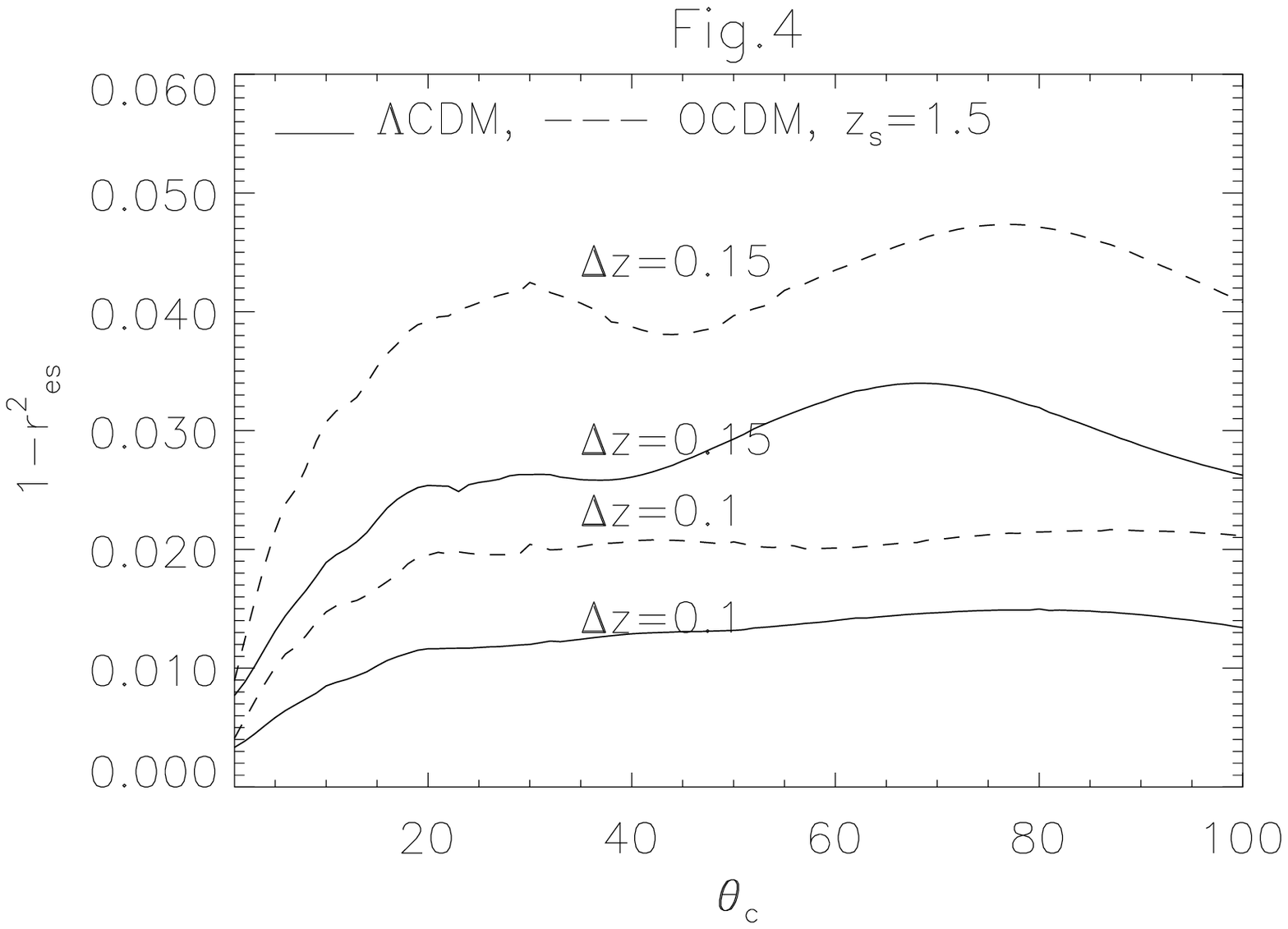}
\caption{Intrinsic uncertainty $1-r_{es}^2$ vs. the angular scale $\theta_c$ (arcmin) with
$z_s=1.5$. The solid lines are for the $\Lambda$CDM model, and the dashed lines are for
the OCDM model. In each pair of lines, the upper one is for $\Delta z=0.15$, and the lower
one is for $\Delta z=0.1$.\label{f4}}
\end{figure}

\begin{figure}
\plotone{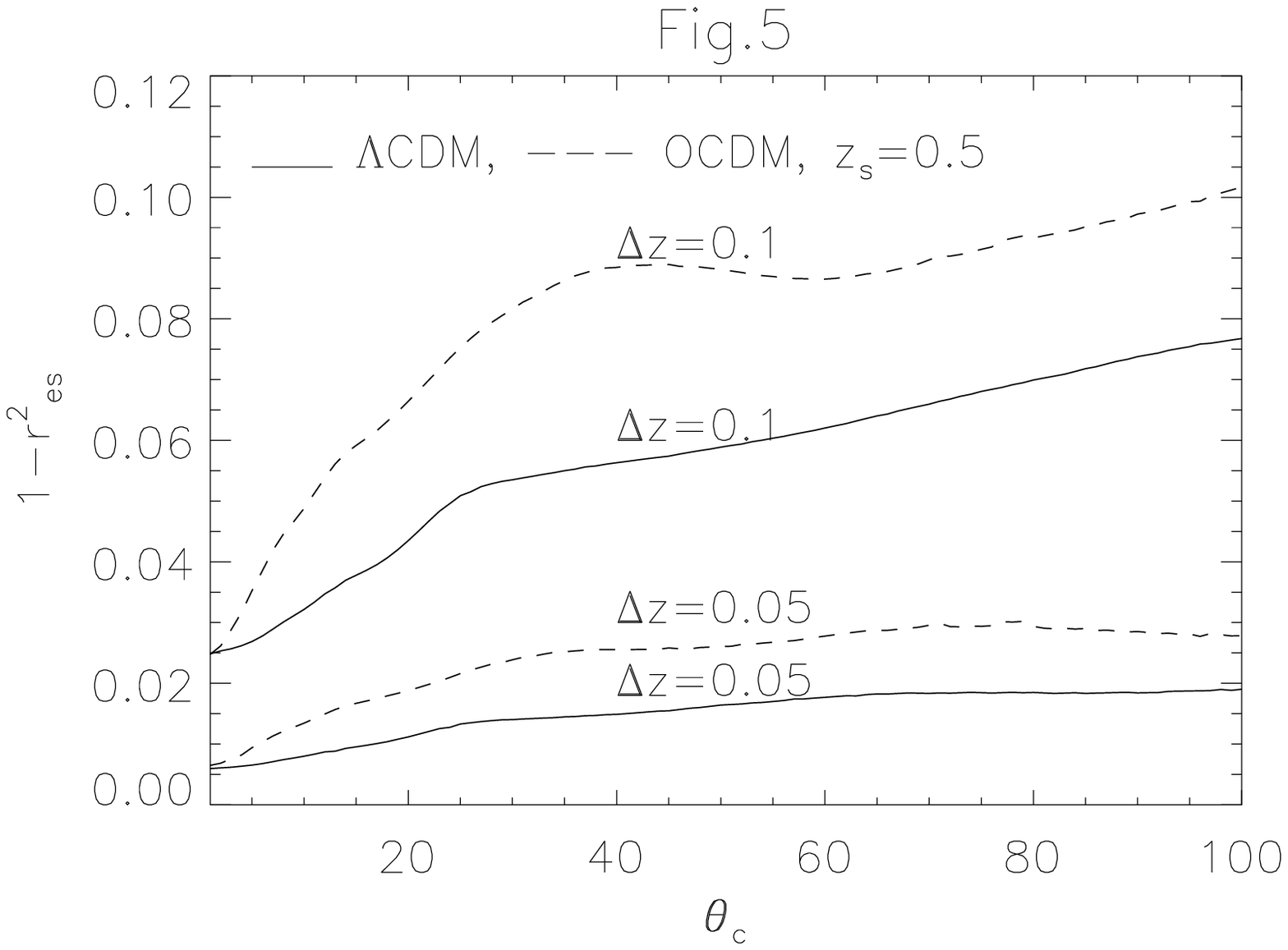}
\caption{Intrinsic uncertainty $1-r_{es}^2$ vs. the angular scale $\theta_c$ (arcmin) with
$z_s=0.5$. The solid lines are for the $\Lambda$CDM model, and the dashed lines are for
the OCDM model. In each pair of lines, the upper one is for $\Delta z=0.1$, and the lower
one is for $\Delta z=0.05$.\label{f5}}
\end{figure}

\begin{figure}
\plotone{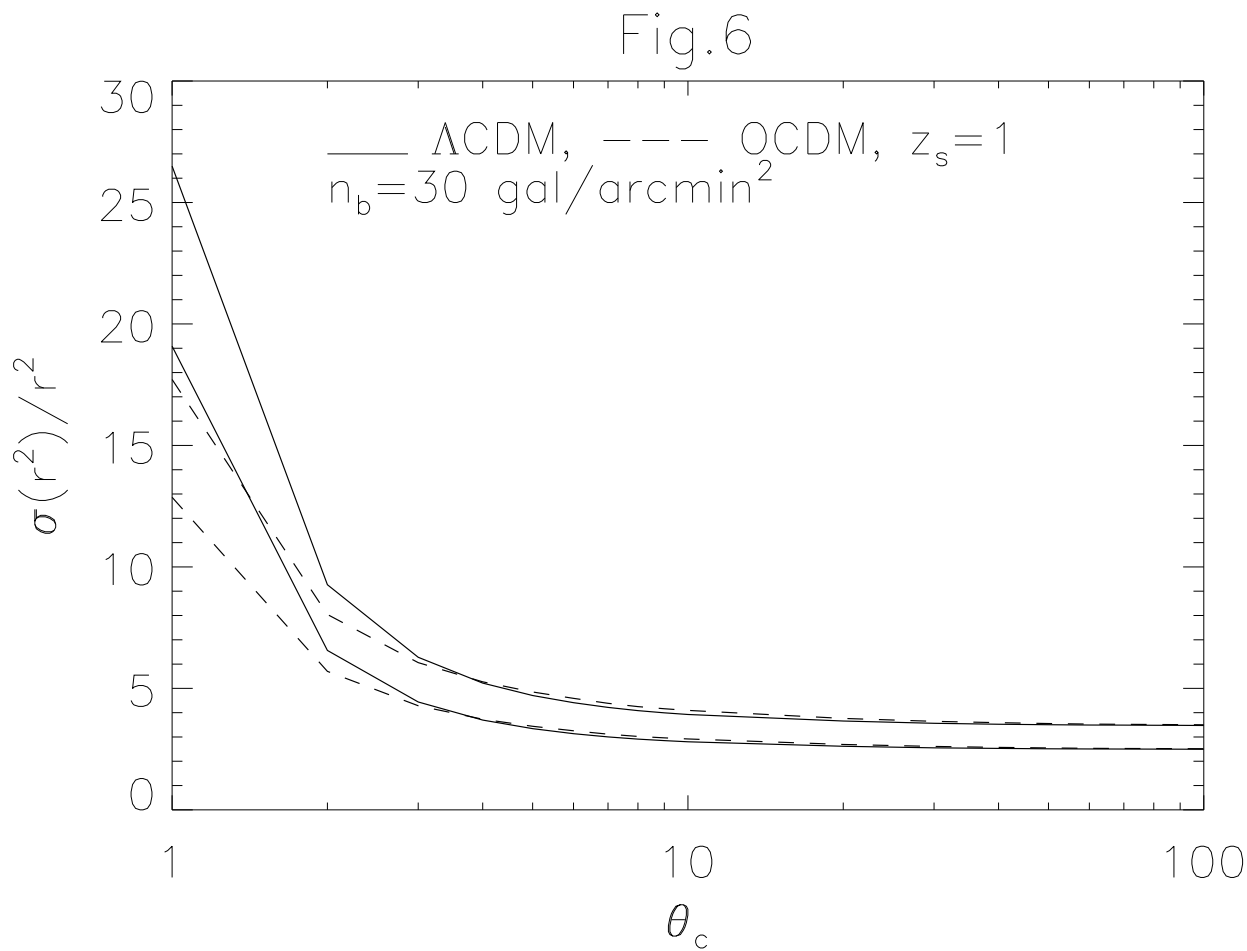}
\caption{Plot of $\sigma(r^2)/r^2$ vs. $\theta_c$ (arcmin) for one field with $z_s=1$,
$n_b=30 \hbox { gal/arcmin}^2$, $n_f=5 \hbox { gal/arcmin}^2$, and $\sigma_{\epsilon}=0.2$.
The solid lines are for the $\Lambda$CDM model, and the dashed lines are for
the OCDM model. In each pair of lines, the upper one corresponds to $\sigma_{max}$ in eq. (23),
and the lower one corresponds to eq. (22).\label{f6}}
\end{figure}

\begin{figure}
\plotone{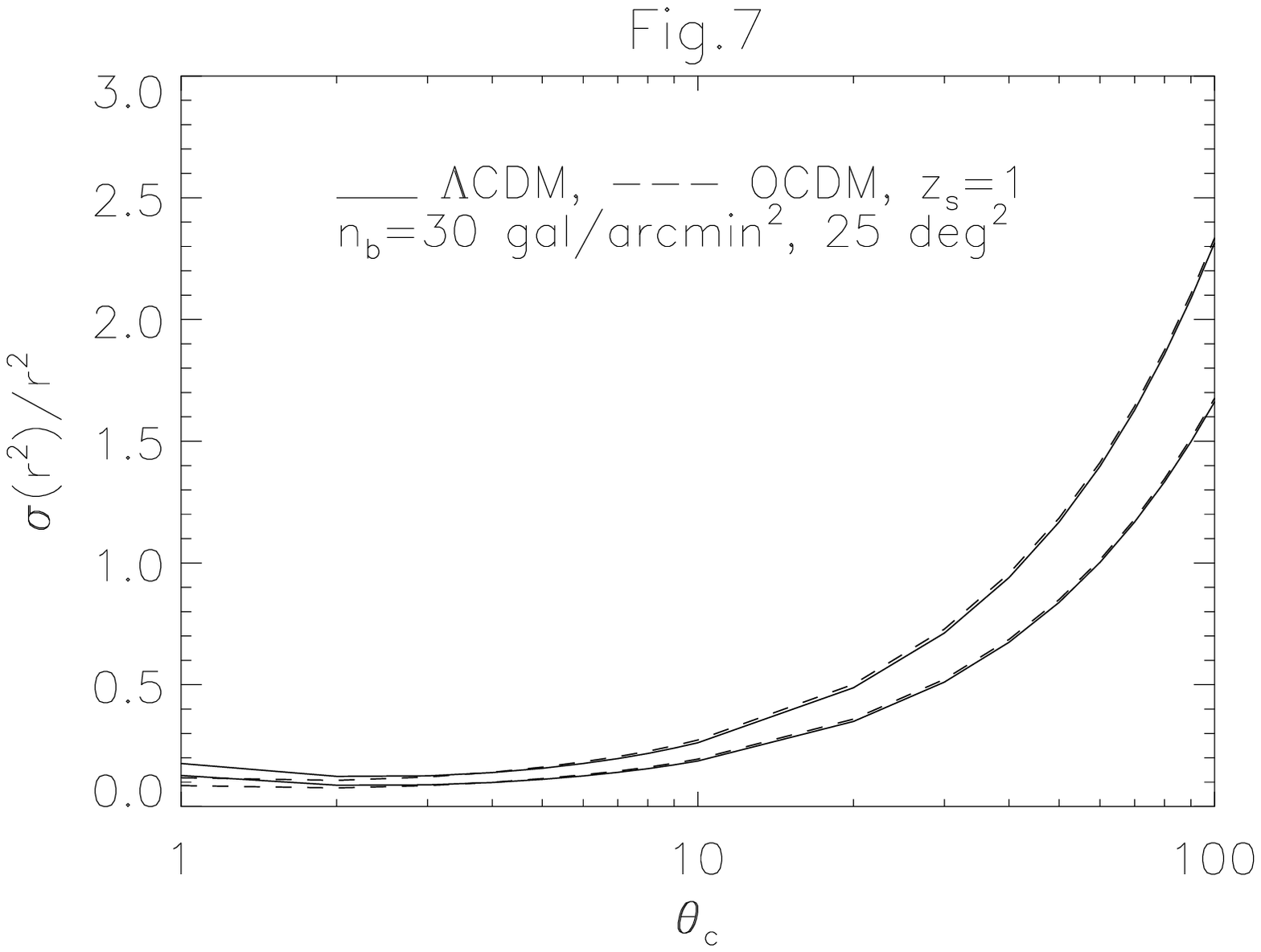}
\caption{Plot of $\sigma(r^2)/r^2$ vs. $\theta_c$ (arcmin) with 
a survey area of $25\hbox { deg}^2$.
All the other parameters are the same as in Fig.6.\label{f7}}
\end{figure}

\begin{figure}
\plotone{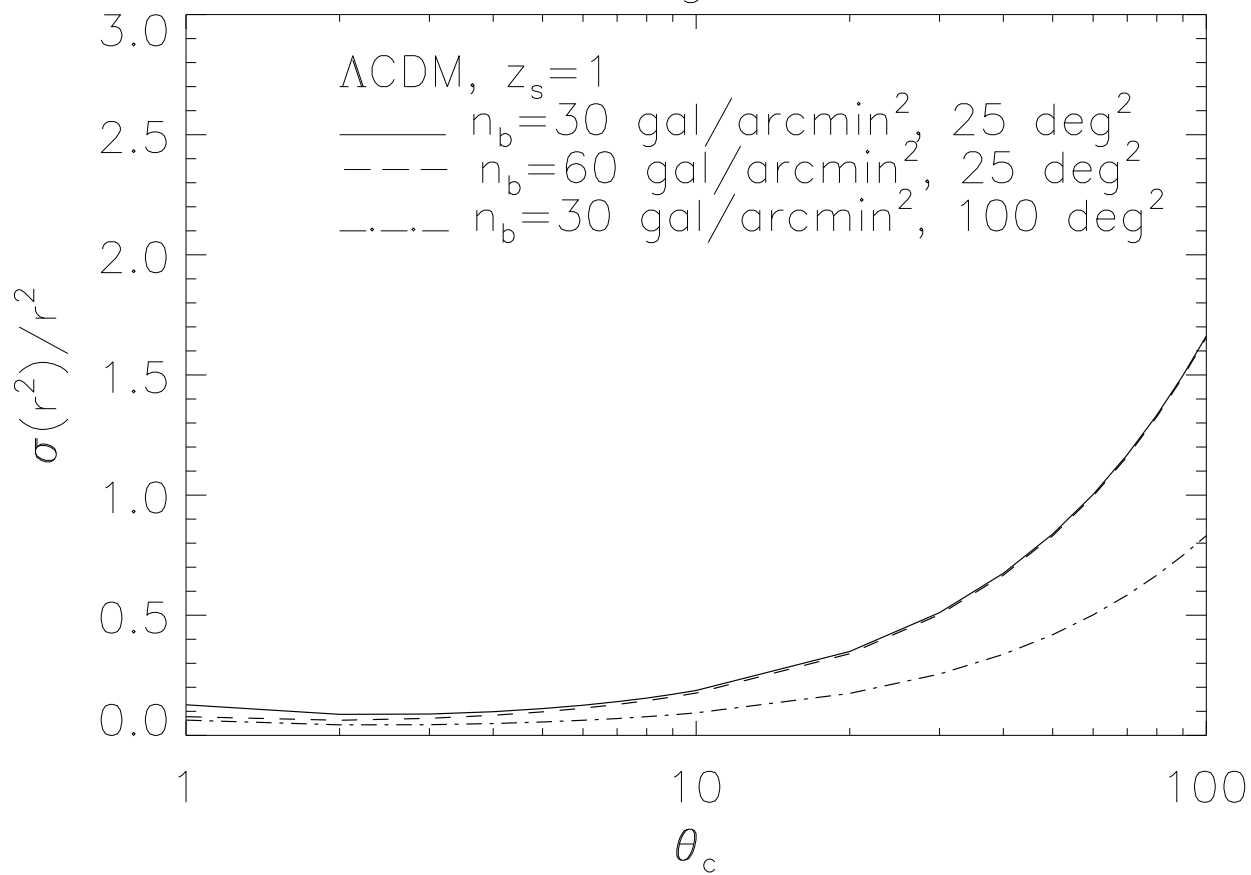}
\caption{Plot of $\sigma(r^2)/r^2$ vs. $\theta_c$ (arcmin) for the 
$\Lambda$CDM model with $z_s=1$ (eq.[22]).
The solid line is for $n_b=30 \hbox { gal/arcmin}^2$ and the survey area of $25\hbox { deg}^2$,
the dashed line is for $n_b=60 \hbox { gal/arcmin}^2$ and a survey area of $25\hbox { deg}^2$,
and the dot-dashed line is for $n_b=30 \hbox { gal/arcmin}^2$ and the survey area of
$100 \hbox { deg}^2$.\label{f8}}
\end{figure}

\begin{figure}
\plotone{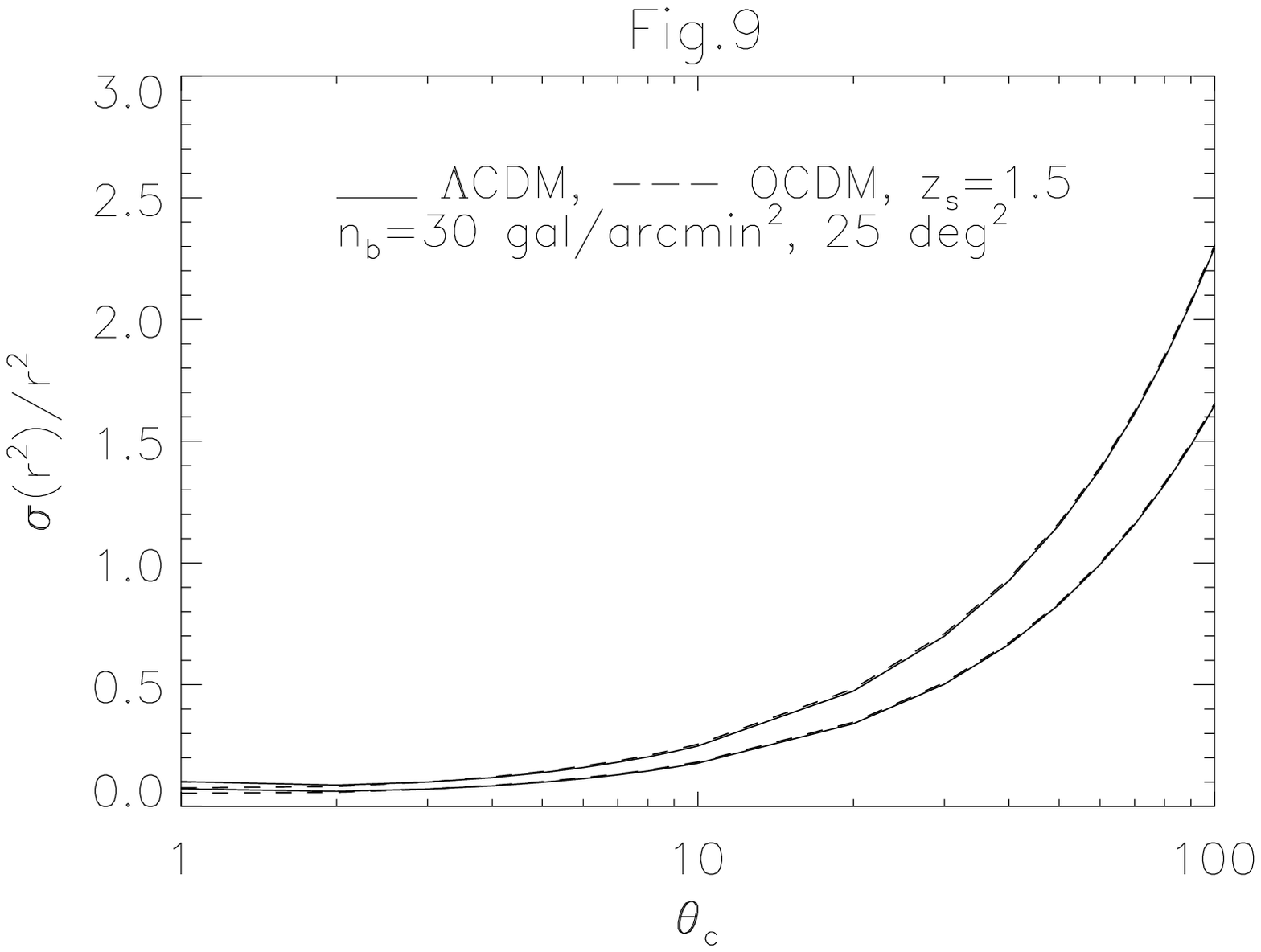}
\caption{Plot of $\sigma(r^2)/r^2$ vs. $\theta_c$ (arcmin) with $z_s=1.5$, 
$n_b=30 \hbox { gal/arcmin}^2$,
$n_f=5 \hbox { gal/arcmin}^2$, and $\sigma_{\epsilon}=0.2$. and a survey area of
$25\hbox { deg}^2$. The meanings of the lines are the same as in Fig.6.\label{f9}}
\end{figure}

\begin{figure}
\plotone{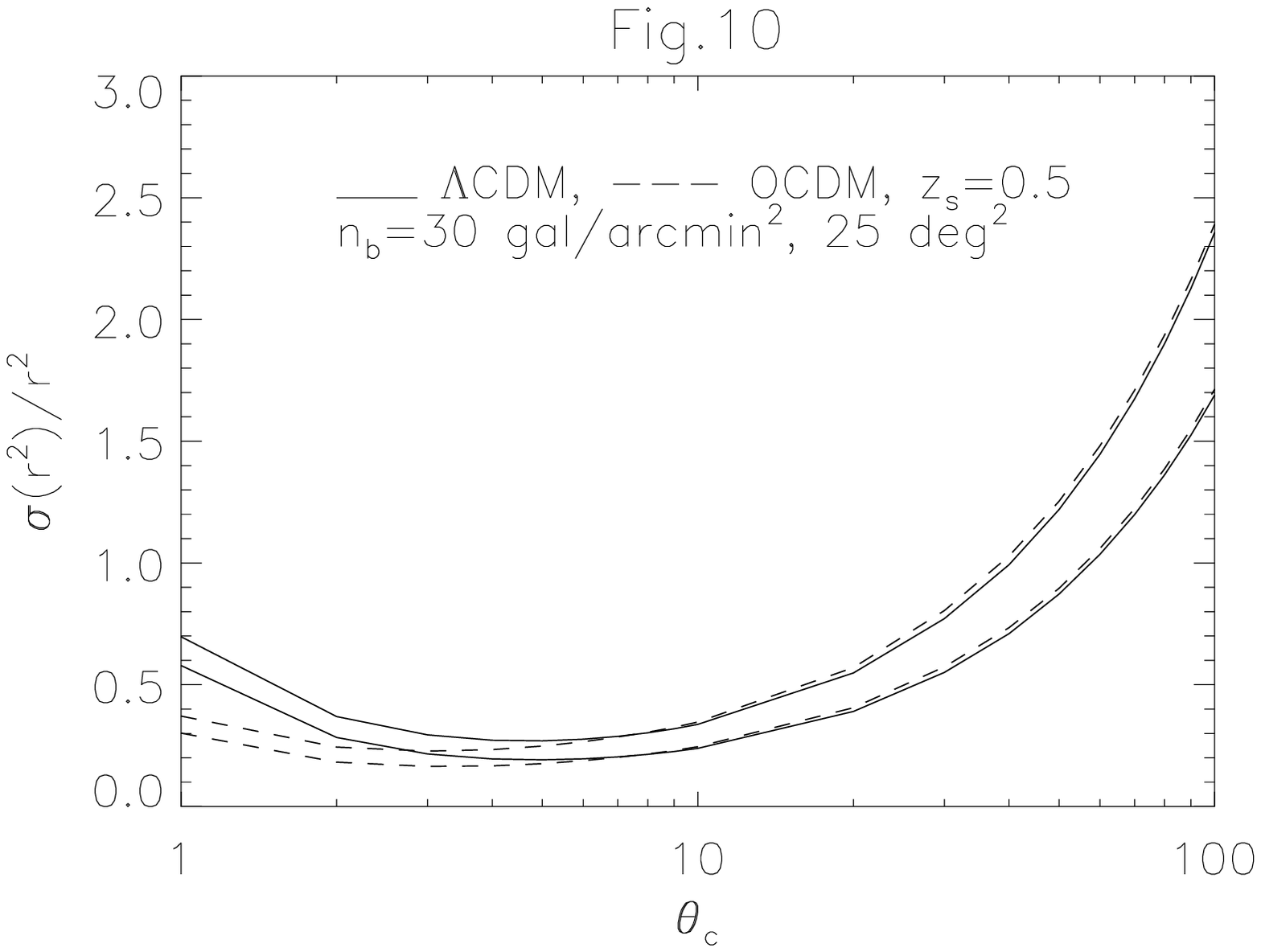}
\caption{Plot of $\sigma(r^2)/r^2$ vs. $\theta_c$ (arcmin) with $z_s=0.5$, 
$n_b=30 \hbox { gal/arcmin}^2$,
$n_f=5 \hbox { gal/arcmin}^2$, $\sigma_{\epsilon}=0.2$, and a survey area of
$25\hbox { deg}^2$. The meanings of the lines are the same as in Fig.6.\label{f10}}
\end{figure}

\begin{figure}
\plotone{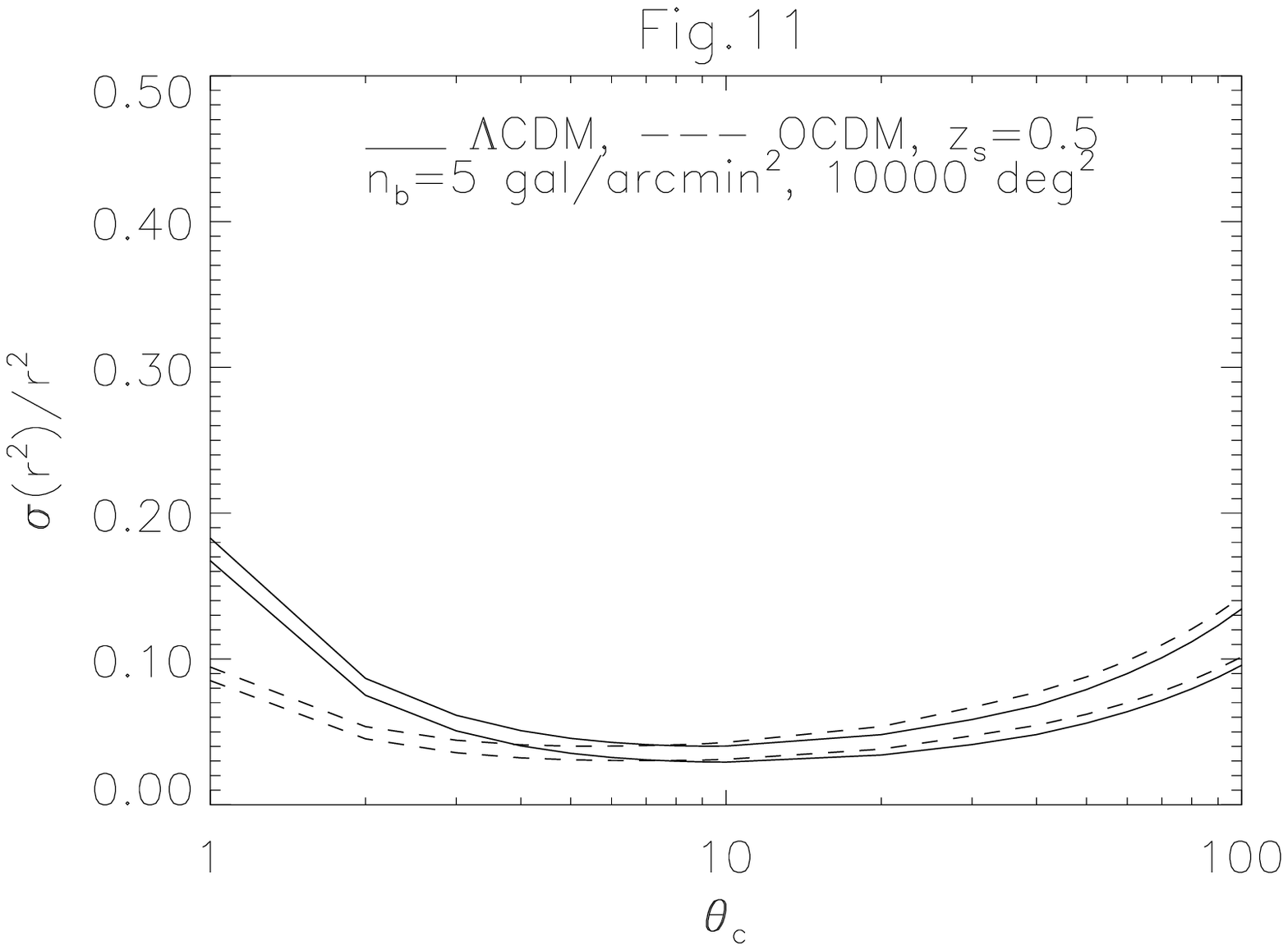}
\caption{Plot of $\sigma(r^2)/r^2$ vs. $\theta_c$ (arcmin) with $z_s=0.5$, 
$n_b=5 \hbox { gal/arcmin}^2$,
$n_f=5 \hbox { gal/arcmin}^2$, $\sigma_{\epsilon}=0.2$, and a survey area of
$10^4\hbox { deg}^2$. The meanings of the lines are the same as in Fig.6.\label{f11}}
\end{figure}

\clearpage






\end{document}